\documentclass[aps,prl,twocolumn,superscriptaddress]{revtex4-1}
\usepackage{bm}
\usepackage{amssymb}
\usepackage{amsmath}
\usepackage{graphicx}
\usepackage{color,soul}
\usepackage{threeparttable}

\begin{document}
\title{Laser Resonance Chromatography of Superheavy Elements}
\author{Mustapha~Laatiaoui }
\email{mlaatiao@uni-mainz.de}
\affiliation{Department Chemie, Johannes
Gutenberg-Universit\"at, Fritz-Strassmann Weg 2, 55128 Mainz,
Germany} \affiliation{Helmholtz-Institut Mainz, Staudingerweg 18,
55128 Mainz, Germany} \affiliation{GSI
Helmholtzzentrum f\"ur Schwerionenforschung GmbH, Planckstrasse 1,
D-64291 Darmstadt, Germany} \affiliation{KU Leuven, Instituut voor
Kern- en Stralingsfysica, Celestijnenlaan 200D, B-3001 Leuven,
Belgium}
\author{Alexei~A.~Buchachenko}
\affiliation{CEST, Skolkovo Institute of Science and Technology, Skolkovo
Innovation Center, Nobel Street 3, Moscow 121205, Russia}
\affiliation{Institute of Problems of Chemical Physics RAS,
Chernogolovka, Moscow District 142432, Russia}
\author{Larry~A.~Viehland }
\affiliation{Science Department, Chatham University, Pittsburgh,
Pennsylvania 15232, USA}
\date{\today}

\begin{abstract}
Optical spectroscopy constitutes the historical path to accumulate
basic knowledge on the atom and its structure. Former work based on
fluorescence and resonance ionization spectroscopy enabled
identifying optical spectral lines up to element 102, nobelium. The
new challenges faced in this research field are the refractory
nature of the heavier elements and the decreasing
production yields. A new concept of ion-mobility-assisted
laser spectroscopy is proposed to overcome the sensitivity limits of
atomic structure investigations persisting in the region of the
superheavy elements. The concept offers capabilities of both
broadband-level searches and high-resolution hyperfine spectroscopy
of synthetic elements beyond nobelium.
\end{abstract}
\pacs{}
%\keywords{}
\maketitle

Chemical elements exhibit atomic emission spectra that are unique
and serve as fingerprints. These spectra have long been observed
for most of the elements, in the laboratory and even from stars, and
have been key ingredients in understanding the cosmic origin of
matter~\cite{Frebel:2018}. Their exploration to ever higher accuracy
in ever heavier elements provides a fertile ground to advance our
understanding of the atom's structure, aside from the classic
advantage of bridging atomic and nuclear physics~\cite{Blaum:2013}.
Precise spectroscopic measurements of hyperfine structures and
isotope shifts of spectral lines enable the study of single-particle and
collective properties of atomic nuclei such as nuclear spins,
moments, and mean square charge radii and provide anchor points for
nuclear models that are presently applied to pinpoint the ``island
of stability'' of superheavy elements.

A great leap forward in this research field was recently achieved
with successful laser spectroscopy of the element
nobelium~\cite{Laatiaoui:2016}. The experiments provided powerful
benchmarks for atomic model calculations~\cite{Chhetri:2018} and
enabled information on the atomic nucleus to be obtained independently
from nuclear models~\cite{Raeder:2018}. Beyond nobelium, only
predictions of the atomic structure exist, which in general are far
from sufficient to reliably identify atoms from spectral
lines~\cite{Dzuba:2017}.

Experiments remain the only means to discover atomic
lines of the heaviest elements and to extend the reference data for
\textit{ab initio} calculations. However, the lack of primordial
isotopes of these elements combined with the inability of breeding
macroscopic amounts in high-flux nuclear reactors renders impossible
the traditional way of studying their atomic emission spectra with
light from primed arc-discharge tubes~\cite{Worden:2008}. The
various techniques currently used to study exotic atoms rely on (i)
the atomic structure already being experimentally known, which is
not the case for elements beyond nobelium, and (ii)
low-energetic atomic beams of the highest quality,
from, e.g., Isotope Separator On-Line facilities~\cite{VanDuppen:2006} that are unsuited to superheavy
element production.

Superheavy elements are produced, at best, at rates less than one
atom per second in heavy-ion induced fusion-evaporation reactions
utilizing powerful particle accelerators in conjunction with
thin-target production techniques~\cite{Schaedel:2002}.
Recoil separators are used to
separate the reaction products from the intense primary beam. The
highly charged product ions may have kinetic energies up to
multiples of ten MeV and have to first be slowed down to thermal
energies for optical spectroscopy. This is best done by stopping
them in gas catchers, preferably in inert gases like helium (He) to
avoid formation of chemical compounds~\cite{Backe:2015}. The ions
are thermalized typically in $1+$ and $2+$
states~\cite{Droese:2013,Lautenschlaeger:2016,Droese:2014,Kaleja:2019}.
Thus far, it has been necessary to neutralize them for resonance
ionization spectroscopy (RIS) as in the nobelium
studies~\cite{Laatiaoui:2016}. In these experiments, electric fields
are used to capture the ions on a filament where neutralization
takes place. By pulse heating the filament, the accumulated fusion
products are released as neutral atoms for subsequent RIS
measurements.

Since this release process depends on the surface material and on
the physicochemical properties of the sample
atoms~\cite{Laatiaoui:2014,Murboeck:2020}, it may result in the emission of
unwanted surface ions at elevated temperatures and thus
substantially hamper the applicability of such a RIS method to the
refractory transition metals in the region of the superheavy
elements. Large efforts are currently undertaken to
find ways for a quick neutralization of the product ions in
typically large buffer-gas stopping volumes without utilizing
catcher filaments, such as exploiting strong beta
sources~\cite{Ferrer:2013}, but a solution for RIS-based methods
cannot yet be safely predicted. Furthermore, any RIS on superheavy
elements would require an extensive search for expected optical
transitions within a broad spectral range inferred from atomic model
calculations. This in turn necessitates the usage
of broadband lasers and aims at simple two-step photoionization
schemes~\cite{Letokhov:1987}. But since ions exhibit ionization
potentials that are energetically far above what could be accessible
utilizing advanced optical setups, this approach remains limited to
neutralized fusion products. Contemporary methods
based on fluorescence detection lack sensitivity due to the
typically limited solid-angle coverage of detectors in the presence
of elevated backgrounds of various origins~\cite{Campbell:2016} and
thus are not suited for such studies.
\begin{figure}[t]
\includegraphics[scale=0.52]{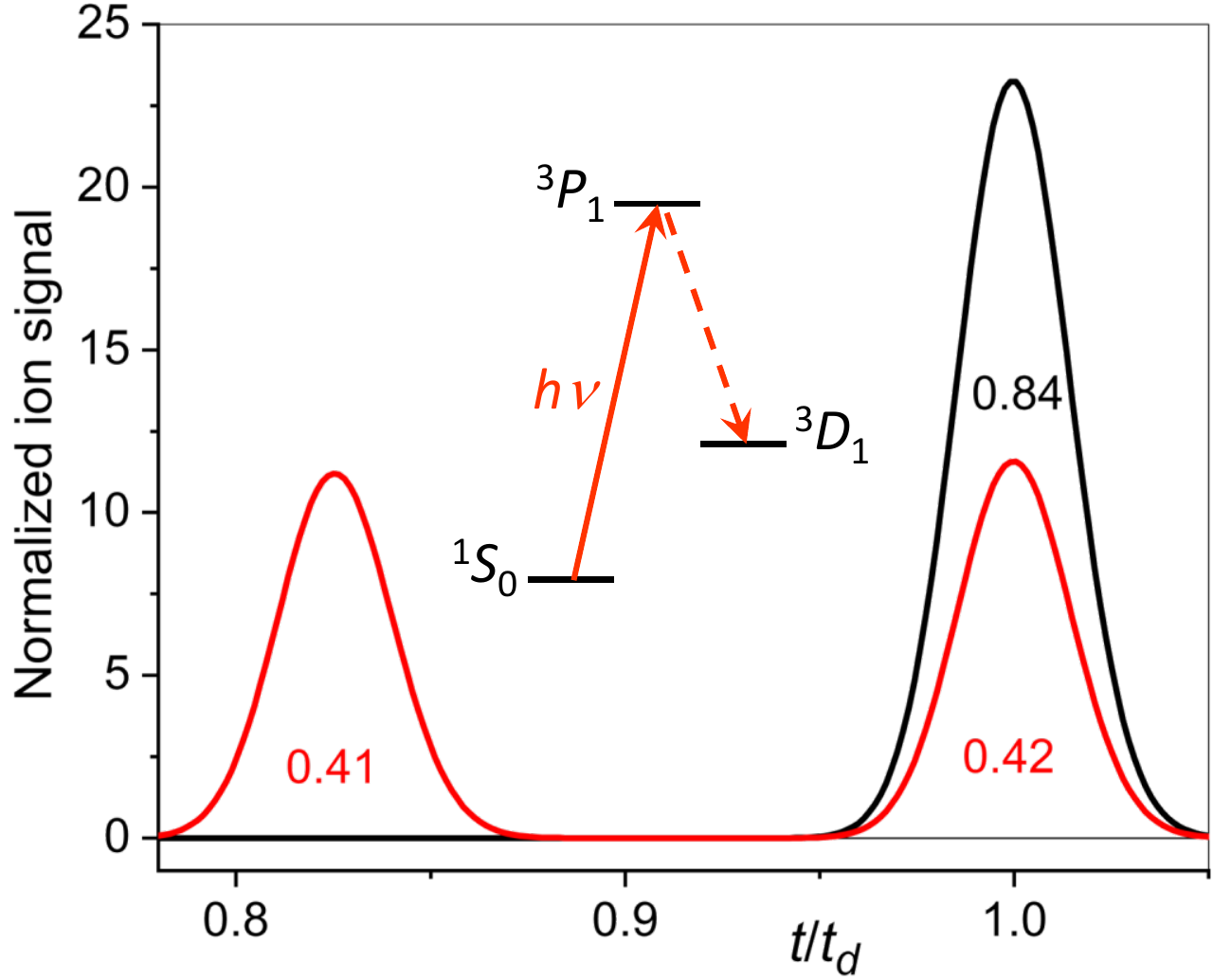}
\caption{\label{fig_ATD}
Lu$^+$ off-resonance (black curve) and resonant
pumping (red curve) signals expected at an effective
temperature of metastable ions of $880\,$K. $L=4\,$cm,
$p_0=3.5\,$mbar He, $T=100\,$K, $E/n_0=38\,$Td, $t_d=25\,\mu$s.
For the resonant pumping signal, $94$\% of the ions have an
initially occupied $^3D_1$ state. Numbers within the peaks
indicate the respective collection efficiencies. (Inset) Simplified
scheme for initial optical pumping by laser radiation ($h\nu$)
from the $^1S_0$ to $^3D_1$ state.}
\end{figure}

In this Letter, an alternate way of optical spectroscopy---termed
laser resonance chromatography (LRC)---is proposed for probing the
heaviest product ions \textit{in situ}, as extracted from gas catchers behind recoil
separators. Its key idea is to detect the products
of resonant optical excitations by their characteristic drift time
to a particle detector. The method exploits laser excitation of the
extracted ions from the ground state to a bright intermediate level
(see the inset in Fig.~\ref{fig_ATD}). An efficient relaxation to
lower-lying metastable states is expected to occur via radiative
processes, which is typical for transition metal ions. Since ion
mobilities are sensitive to changes in electronic
configurations~\cite{Laatiaoui:2012,Viehland:2018}, such as during
optical pumping, guiding these ions through a drift tube enables one
to discriminate ions in ground and metastable states by drift times.
This effect of electronic state chromatography is well established
from ion-mobility spectrometry of many transition
metals~\cite{Kemper:1991,Taylor:1999,Iceman:2007,Ibrahim:2008,Armentrout:2011,Manard:2016,Manard:2016b}.
The time spectra obtained without resonant excitations characterize
initial background ions, whereas detecting ions at significantly
shorter or longer times signals resonant pumping, which triggers
the conversion of the background ions into the metastable ones. LRC
has a number of key advantages compared to common optical
spectroscopy. Omitting intermediate steps such as neutralization of
thermalized fusion products, photoionization of ions, and radiation
detection provides a general optical access to
elemental cations without sacrificing speed or
sensitivity.

The method can be applied to element 103, lawrencium (Lr), which is
currently the heaviest element in the focus of RIS
experiments~\cite{Ferrer:2017,Laatiaoui:2019}. As follows from
quantitative \textit{ab initio} calculations for
Lr$^+$~\cite{Kahl:2019}, pumping the ground state $7s^2\,^1S_0$ to
the bright $7s7p\,^3P_1$ state effectively feeds the lowest
$6d7s\,^3D_1$ state of a radiative lifetime ($25$ days) exceeding
the half-lives of all known Lr isotopes. The
efficiency for optical pumping depends on both the strength of the
ground-state transition to the bright intermediate level and on the
branching ratios from this level to the lower-lying ground and
metastable states. Calculations predict a sizable Einstein
coefficient for spontaneous emission of
$A_{ki}=6.36\times10^7\,S^{-1}$ for the $^1S_0$-$^3P_1$
ground-state transition~\cite{Kahl:2019}. The intermediate $^3P_1$
level is predicted at $31,540\pm389\,$cm$^{-1}$ (about $317\,$nm),
which is readily accessible for laser probing. The branching ratio
to lower-lying metastable states is expected to be $10$\% for
Lr$^+$, but it can exceed $50$\% as in the lighter homologue
Lu$^+$~\cite{Quinet:1999,Paez:2016}. We implemented a five-level
system into a rate equation model to calculate the probability for
optical pumping of the metastable $^3D$ states in Lu$^+$ and
Lr$^+$~\cite{Laatiaoui:2020b}. In the model we assumed a laser-ion
interaction of $10\,$ns duration every $100\,\mu$s at an energy
density of the laser radiation of $10\,\mu$J/cm$^2$. In addition, we
accounted for instantaneous population transfers due to possible ion
collisions with residual gas by including the same state-quenching
rates as those reported for the isoelectronic neutral barium in
He~\cite{Brust:1995}. Starting with a level system in its
ground state ($^1S_0$), we obtained a $31.5$\% probability for
optical pumping into the $^3D_1$ state in Lu$^+$ already for one
single pulse exposure. This probability amounts to
$7.5$\% in Lr$^+$ and reaches values of $94$\% and $53$\% after ten
pulses for the two ionic species, respectively, which are
sufficiently high for the LRC studies.
\begin{figure*}
\includegraphics[width=0.74\textwidth]{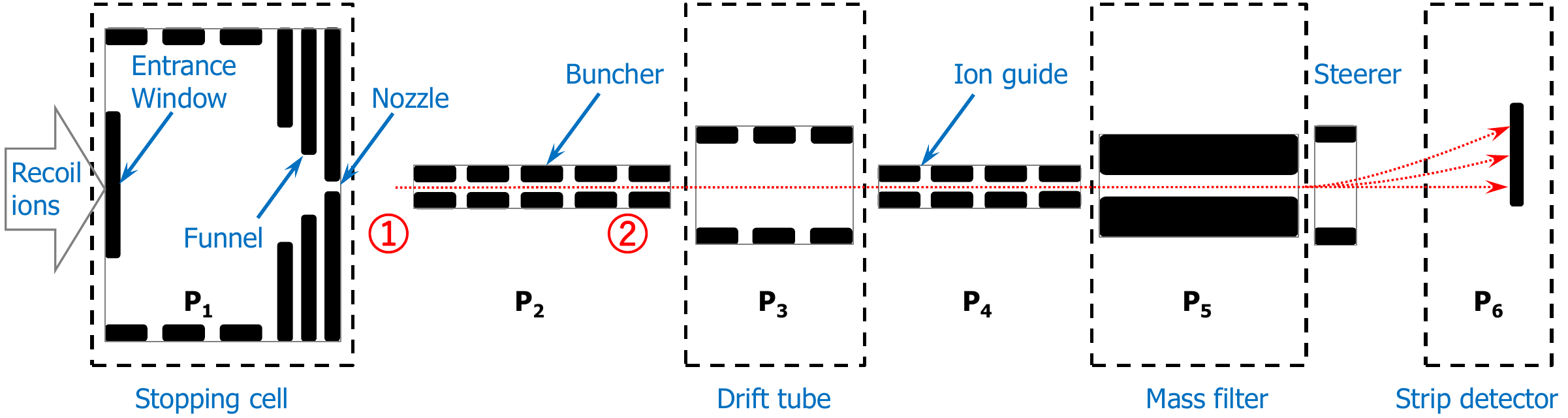}
\caption{Schematic view of the LRC setup for on-line experiments.
Laser probing of ionic states occurs inside the buncher (position 2)
during the level search or in the gas jet (position 1) to enable
hyperfine spectroscopy. The dashed lines show
potential trajectories of the extracted ions along
the apparatus. P$_1\approx60\,$mbar,
P$_2\approx5\times10^{-2}\,$mbar, P$_3=p_0<10\,$mbar,
P$_4\approx10^{-3}\,$mbar, P$_5\approx10^{-5}\,$mbar, and
P$_6\approx10^{-7}\,$mbar. See the text for more details.}
\label{fig_setup}
\end{figure*}

The mobility $K$ is known for many atomic species in many neutral
gases~\cite{LXCAT:2019}. Macroscopically, it relates the speed $v_d$
of an ion drifting in trace amounts through a dilute gas of pressure
$p_0$ and temperature $T$ to the applied electric field strength $E$
by the equation $v_d=K\,E$~\cite{Mason:1988}. Usually, it is reported
as the reduced mobility, $K_0=K$($p_0$/$p_{st}$)($T_{st}$/$T$), which
implies normalization to the standard temperature ($T_{st}=273.15\,$K)
and pressure ($p_{st}=1013.25\,$mbar). The mobility can be derived,
together with the diffusion coefficients, from the solution of the
Boltzmann equation for a steady-state drift of an ion
swarm~\cite{Mason:1988,Viehland:1994,Viehland:2018}. Momentum
transfer and similar transport cross sections of the ion-neutral
collisions are the main parameters that establish the microscopic
relation between the transport coefficients, e.g.,
mobility, and the ion-neutral interaction potential.
As the repulsive short-range interaction depends on
the electronic configuration, the mobility inevitably inherits this
dependence, underpinning the electronic-state-chromatography
effect.

Accurate measurements for transition
metals~\cite{Taylor:1999,Iceman:2007,Ibrahim:2008} revealed that the
mobility difference may be as large as $33$\% in the case of
different states of Cu$^+$ ions~\cite{Ibrahim:2008}.
Recent systematic measurements across the
lanthanides~\cite{Manard:2017} have shown, in agreement with former
predictions~\cite{Buchachenko:2014}, that the zero-field
mobilities of ions with the $4f^n6s$ configuration in He deviate from
$19.0\,$cm$^2$/Vs within $\pm3$\%, whereas for Gd$^+$ ($4f^75d6s$) and
Lu$^+$ ($4f^{14}6s^2$) ions the deviations approach $+9$\% and $-12$\%,
respectively.

To verify the perspectives of discriminating Lr$^+$ ions in their
ground and metastable states, we performed extensive calculations
for Lu$^+$, which has a similar structure of the excited state but,
unlike its heavier homologue, has been studied
experimentally~\cite{Manard:2017}. The interaction potential for
Lu$^+$ in its ground state with He was computed in
Ref.~\cite{Buchachenko:2014} using the single-reference coupled
cluster method with singles, doubles, and noniterative correction to
triples. The standard mobility in He at room
temperature was predicted in these calculations as $16.6\,$cm$^2$/Vs
and was found to be accurate within $2$\% when compared with the
more recent value of $16.8\pm0.4\,$cm$^2$/Vs reported from mobility
measurements at $295\,$K~\cite{Manard:2017}. A similar approach was
followed here to address the interaction of Lu$^+$ in the
$6s5d$~$^3D_1$ metastable state~\cite{Laatiaoui:2020b}.
Configuration interaction methods~\cite{Berning:2000} were
implemented to account for spin-orbit coupling.
Transport coefficients were computed by solving the Boltzmann equation
according to the Gram-Charlier approach~\cite{Viehland:1994,Viehland:2018}.
Momentum transfer cross sections for the $\Omega =
0^-$ and $1$ components of the metastable $^3D_1$ state were
averaged to give total cross sections in a way verified recently for
the open-shell Gd$^+$~\cite{Buchachenko:2019} for which a
$1$\% relative accuracy was achieved. Thus, we expect a similar
accuracy for mobilities of Lu$^+$ in the $^3D_1$ state as for this
ion in the ground state.

The calculated transport coefficients indicate that
Lu$^+$ ions in the metastable state drift faster than the ones in the
ground state but experience greater diffusion~\cite{Laatiaoui:2020b}.
In addition, the deviation in ion mobility can be as high as $24$\%.

To better assess the LRC separation capabilities for Lu$^+$ in
the different states, we used the analytical expression derived in
Ref.~\cite{Moseley:1969} for describing ion swarms drifting in
dilute gases~\cite{Laatiaoui:2020b}. We took an ion
swarm starting in the initial state occupation as obtained from the
rate equation model after ten laser-pulse exposures and calculated
the drift time distributions throughout a drift length $L$ for
different pressures, temperatures, and electric fields to gas number
densities $E/n_0$~\cite{Laatiaoui:2020b}.
Figure~\ref{fig_ATD} shows these distributions for a drift
length of $4\,$cm, $T=100\,$K, $E/n_0=38\,$Td, with $1\,$Td (Townsend unit) being equal to $10^{-21}\,$Vm$^2$, and $p_0=3.5\,$mbar,
which were found to be a good compromise between sufficient time
resolution and detection efficiency. In this case, the mean drift
time of the ground state ions is $25\,\mu$s. Importantly, we show
that about $41$\% of the ions can be collected in the $^3D_1$ state
after pumping and drift, despite collisional quenching to the ground
state at a rate as measured for barium in He~\cite{Brust:1995} at $880\,$K.
Applying smaller $E/n_0$ ratios results in less ion heating, which
could be necessary to suppress quenching to the ground state at the
cost of diffusion losses. Still, a collection efficiency above $28$\%
can be maintained for effective ion temperatures up to $300\,$K
under optimal conditions when longer drift times are envisaged. Also
reducing the initial $^3D_1$ occupation to $53$\%, as expected
from optical pumping in Lr$^+$, results in sufficiently high
collection efficiencies above $16$\%~\cite{Laatiaoui:2020b}.

From the practical viewpoint, we propose a LRC setup as shown in
Fig.~\ref{fig_setup}. It consists of a buffer-gas stopping cell and
a detection part allowing for cooling, guiding, selecting, and
detecting the ions under investigation. We suggest using helium as a
buffer gas in the stopping cell at pressures around $60\,$mbar,
following similar gas catcher concepts devoted to mass
measurements~\cite{Neumayr:2006,Block:2010}. The cell incorporates a
radio-frequency (rf) funnel and is isolated from the detection part
by an extraction nozzle~\cite{Droese:2014,Kaleja:2019}. The
detection part comprises, sequentially, a rf quadrupole in a
bunch-mode operation (called a buncher), a cryogenic drift tube, a
second rf quadrupole as an ion guide, a quadrupole
mass spectrometer (QMS) for charge-to-mass selection, and a
particle detector. Our approach stipulates the use of very short
drift tubes filled with He at pressures $<10\,$mbar
in order to suppress unwanted quenching and to pursue the
desired resolving power by applying appropriate gas temperatures and
$E/n_0$ ratios. Similar drift tube concepts have been followed in
state-specific ion-mobility studies along the third-row transition
metals~\cite{Iceman:2007,Ibrahim:2008}.

The fusion products provided by recoil separators penetrate the
stopping cell through an entrance foil and thermalize mostly as ions
in high purity helium~\cite{Minaya:2012}. Adding
small quantities of chemically inert gases like krypton or xenon of
relatively low ionization thresholds into the buffer gas helps to
shift the ion charge states to $1+$ without significant ion
losses. The ion manipulation by electric dc and rf fields enables a
fast and efficient extraction of the fusion products within the
emerging buffer gas that flows continuously through the nozzle.
Using a laser beam of tunable wavelengths from a pulsed laser
system~\cite{Kudryavtsev:2013}, one can search for ground-state
transitions of the sample ions trapped in the
buncher~\cite{Toschek:1980} (position 2 in Fig.~\ref{fig_setup}) at
background pressures $\leq5\times10^{-2}\,$mbar. The expected
Doppler and power broadening of the spectral lines enable a
relatively quick and efficient search for optical resonances. The
singly charged ions will then be in either the ground- or the
metastable-state configuration during their transport downstream to
the drift tube. Gas cooling inside the buncher will facilitate the
injection into the drift tube. Bunched ions passing the drift tube,
the ion guide, and the QMS will be detected with, e.g., a
Channeltron detector placed in a high-vacuum section.
The time differences between buncher timing and
detector-signal time stamps describe arrival time distributions,
which will be monitored for long time periods during the level
search. At constant background conditions, a high sensitivity for
resonance detection can be reached because successful laser
excitations trigger optical pumping, change the ratio of metastable ions
to ground-state ions of distinct mobilities, and thus cause a change
in the time distributions~\cite{Laatiaoui:2020b}. On establishing a
ground-state transition, like the predicted $^1S_0$-$^3P_1$
transition in Lr$^+$, much more precise LRC measurements can be
performed by probing the ions transported in the
cold supersonic gas jet of low
density~\cite{Ferrer:2017} formed in front of the nozzle (position 1
in Fig.~\ref{fig_setup}). In the case of elevated background ions,
an increased sensitivity of the setup can be achieved by a
synchronized steering of the mass-selected ions along a
position-sensitive Si-strip detector (SSD). Since
the radionuclides of interest are $\alpha$ emitters, the drift time
information will then be imprinted in the position of the
$\alpha$-activity hot spots on the Si detector
(Fig.~\ref{fig_setup}). This way of ion detection allows (i)
a zoom in the time line when the steering process
is initiated at the expected arrival time of the ions and (ii)
a nearly background-free ion detection at rates
down to one count per hour or even less due to the fingerprint
$\alpha$-decay detection.

The stopping and extraction of the fusion products can be as
efficient as $10$\% at room temperature~\cite{Block:2015}. An
increased efficiency of about $33$\% can be achieved if a cryogenic
stopping cell is used~\cite{Kaleja:2019}. We expect $50$\% to $70$\%
of the extracted products to be singly charged monoatomic ions, out
of which we expect $50$\% at least to be in their ground state.
Here we give a rather conservative estimate for
this ion fraction, as elevated pressures inside the stopping cell
promote fast relaxation processes during a comparably slow
extraction. In addition, we expect a transmission through the rf quadrupole structures to be about $80$\% for light ions~\cite{Ferrer:2013},
while a slightly higher transmission efficiency ($90$\%) can be
achieved for the heaviest elements. We do not expect losses due to
space charge effects to play a significant role since these usually
alter the transmission for bunches containing more than $10^4$
ions~\cite{Ferrer:2013}, which will not be the case
for elements beyond nobelium. The transmission of the ions through
the QMS amounts to $90$\% at a moderate mass
resolution~\cite{Haettner:2018}. Table~\ref{tab_efficiencies}
summarizes the expected partial efficiencies for the LRC method.
Taking the collection efficiency of the drift tube into account, one
ends up with an overall efficiency of the apparatus of
$0.2$\%-$3.5$\% for Lu$^+$. Similar values
should be reachable for Lr$^+$ and other heavier elements. Although
small, these values are at least $2$ orders of magnitude larger than
what can be achieved for common fluorescence-based spectroscopy and
similar to those recently reported for resonance ionization
spectroscopy across the actinides~\cite{Laatiaoui:2016,
Ferrer:2017}.
\begin{table}
  \caption{Expected partial and overall efficiencies for LRC on Lu$^+$.} \label{tab_efficiencies}
  \centering
  \begin{tabular}{lr}
  \hline\hline
  Type & Efficiency (\%)\\
  \hline
  Stopping and extraction & 10 - 33 \\
  Fraction of $1+$ ions & 50 - 70 \\
  Initial ground-state occupation & 50 \\
  Buncher transmission &  90 \\
  Lu$^+$($^3D_1$) collection from drift tube\footnote{Considers optical pumping and signal losses due to transverse diffusion and quenching~\cite{Laatiaoui:2020b}.} &  28 - 41 \\
  Ion guide transmission &  90 \\
  QMS transmission &  90 \\
  Ion detection\footnote{Taken $100$\% when using a Channeltron and about $40$\% when using a position-sensitive SSD for $\alpha$-decay detection.} & 40 - 100 \\\\
  In total         & 0.2 - 3.5 \\
  \hline\hline
  \end{tabular}
\end{table}

In summary, LRC should be a sensitive and versatile
technique of optical spectroscopy that will enable both broadband-level searches and narrow-band hyperfine spectroscopy on elements thus
far inaccessible in the periodic table beyond nobelium. Atomic model
calculations predict a strong $^1S_0$-$^3P_1$ ground-state
transition in Lr$^+$ at about $317\,$nm, which is envisaged for
future LRC investigations~\cite{Kahl:2019}. In addition, the method
can be used to explore the atomic structure of the superheavy
elements (rutherfordium and dubnium) of refractory nature that are
available only in minute production
quantities~\cite{Block:2015,Ackermann:2015}.

Other ionic species like triply charged thorium
(Th$^{3+}$)~\cite{Campbell:2011,Wense:2016,Wense:2016N} should be
within reach of the LRC method as well. Th$^{3+}$ exhibits a
long-lived metastable atomic state for optical probing as well as
suitable laser cooling transitions~\cite{Safronova:2006} making it
an ideal candidate for the realization of a future frequency
standard based on the nuclear isomer
$^{229m}$Th~\cite{Seiferle:2019N}. Moreover, an axial injection of a
continuous-wave laser beam through the drift tube
makes the method more universal because average drift times of ions
in ground and excited states result whenever the resonance
conditions are fulfilled and the respective
mobilities are distinct. Finally, the ability to optically pump ions
in a controlled way has the potential to make further optical
transitions accessible and to provide state-purified cations for
chemistry experiments~\cite{Armentrout:2011} and benchmark data for
state-of-the-art \emph{ab initio} calculations.

\begin{acknowledgments}
We thank M.~Verlinde, R.~Ferrer, S.~Raeder, F.~Schneider,
T.~Murb\"ock, J.~Berengut, A.~Borschevsky, P.~Van~Duppen, M.~Block,
H.~Backe, and H.-J.~Kluge for the fruitful discussions on the LRC
concept. A.A.B. acknowledges the support of the Russian Foundation for
Basic Research (Project No. 19-03-00144). This project has received
funding from the European Research Council (ERC) under the European
Union's Horizon 2020 research and innovation programme (Grant
Agreement No. 819957).
\end{acknowledgments}

\end{document}